\documentclass[12pt]{article}
\usepackage{amsfonts}
\usepackage{epsfig}
\newcommand{\nn}{\nonumber}

\newcommand{\be}{\begin{equation}}
\newcommand{\eeee}{\end{equation}}
\newcommand{\bd}{\begin{displaymath}}
\newcommand{\ed}{\end{displaymath}}
\newcommand{\bea}{\begin{eqnarray}}
\newcommand{\eea}{\end{eqnarray}}

\renewcommand{\paragraph}[1]{
\vspace{.8mm}\par\noindent {\sl #1}\\
\vspace{0.2mm} }

\newcommand{\ba}{\left(\begin{array}}
\newcommand{\ea}{\end{array}\right)}
\newcommand{\Z}{{\cal Z}}
\newcommand{\mydef}[2]{ \vspace{1.ex}{\bf #1:} #2 \vspace{1.ex} }
\def\R{{\mathbb R}}
\def\C{{\mathbb C}}
\def\Z{{\mathbb Z}}

\def\dim{{\rm dim}\ }
\def\alg{{\mathfrak{alg}}\ }

\begin{document}
\begin{titlepage}
\begin{flushright}
SU-ITP-00/08\\
{\tt hep-th/0002074}\\
\today\\
\end{flushright}
\vskip 2cm

\begin{center}
{\large \textsc{
$p$-Gerbes and Extended Objects in String Theory
}} \vskip 0.7 cm
{\small
{\bf  Yonatan Zunger}\\
\vskip.5cm
{\it Department of Physics, \\
Stanford University, Stanford, CA 94305-4060, USA \\
E-mail: zunger@leland.stanford.edu}
}
\end{center}
\vskip 1cm
\begin{abstract}
$p$-Gerbes are a generalization of bundles that have $(p+2)$-form field
strengths.  We develop their properties and use them to show that 
every theory of $p$-gerbes can be reinterpreted as a gauge theory containing
$p$-dimensional extended objects. In particular, we show that every closed
$(p+2)$-form with integer cohomology is the field strength for a  gerbe,
and that every $p$-gerbe is equivalent to a bundle with connection on the 
space of $p$-dimensional submanifolds of the original space. We also 
show that $p$-gerbes are equivalent to sheaves of $(p-1)$-gerbes, and
use this to define a $K$-theory of gerbes. This $K$-theory classifies
the charges of $(p+1)$-form connections in the same way that bundle $K$-theory
classifies 1-form connections. 
\end{abstract}


\end{titlepage}

\section{Introduction}

$p$-Gerbes are a generalization of fiber bundles which have higher form
connections. For $p=0$, they are bundles. The case $p=1$ was introduced
by Giraud \cite{Giraud} and refined by Brylinski \cite{Brylinski} as a 
tool to study the properties of 3-manifolds. A good introduction to their
properties was given by Hitchin \cite{Hitchin}. The case $p=2$ was
developed in \cite{CareyEtAl} in order to study higher cohomology
classes in gauge theories.

Gerbes are valuable because they provide a geometric way to unify the
properties of $p$-form fields with gauge symmetries. We will begin by
studying the detailed properties of these objects. We will show that
every closed $(p+2)$-form with integral cohomology is the field strength
of a $p$-gerbe, and that $p$-gerbes are equivalent to bundles with 
connection on the space of smooth $p$-dimensional submanifolds of the
original base space. This means that any theory of higher forms implicitly
is a theory of extended objects; at the end of this paper, we will make
this relationship explicit, showing how the higher-form fields can be 
replaced by the integrals of 1-forms over $p$-dimensional internal spaces.

It will also be useful to derive a better topological and geometric 
picture of gerbes. To this end, we study their local properties, and show 
that they have sections; in fact, $p$-gerbes are equivalent to sheaves
of $(p-1)$-gerbes. This allows us to develop a $K$-theory of gerbes
analogous to that of bundles, which (for similar reasons) classifies the
higher-form charges of extended objects in string theory. The results are
consistent with the known NS $B$-field charges in type II string theory.

The paper is laid out as follows. In section 2, we define $p$-gerbes and
introduce three equivalent pictures thereof:

\begin{itemize}
\item{\v{C}ech language: A $p$-gerbe on a manifold $X$ over a Lie group
$K$ can be thought of as an open cover of the space along with $K$-valued
transition functions on $(p+2)$-fold intersections. This language 
contains the underlying definition of a gerbe and is useful for 
computations.}
\item{de Rham language: $p$-gerbes have $(p+2)$-form field strengths and
$(p+1)$-form connections. We show that every closed $(p+2)$-form on a
manifold with integral Chern class is the field strength of some $p$-gerbe.
These gerbes have a gauge symmetry of the form $B\rightarrow B+dA$, where
$B$ is the connection and $A$ is an arbitrary $p$-form. }
\item{Loop language: $p$-gerbes implement gauge symmetries on the space
of $p$-loops in the same way that bundles (which are 0-gerbes) implement
gauge symmetries on the original space. In particular, gauge symmetries
involving closed strings are naturally associated with 1-gerbes, and the
Neveu-Schwarz tensor field can be interpreted as the connection associated
with such a symmetry.}
\end{itemize}

In section three, we study their local structure and define a fourth
picture:
\begin{itemize}
\item{Sheaf language: At least for Abelian $K$, a $p$-gerbe is a sheaf of
$(p-1)$-gerbes. An example (due to Hitchin) is a space which does
not support a Spin$^c$ structure for topological reasons; such a space
can be covered with open sets, on each of which such a structure is defined.
The combination of all such sets and their transition functions forms a
1-gerbe, since each structure is a bundle.\footnote{There is also a fifth 
language, that of sheaves of groupoids, in terms
of which (1-)gerbes were originally introduced; we will not discuss this
here, but refer the interested reader to \cite{Brylinski}.}
}
\end{itemize}

In section 4, we use the \v{C}ech and sheaf pictures to define the $K$-theory
of gerbes, and show that it behaves very similarly to that of bundles. 
Finally, we return to the question of how extended objects emerge from
gerbe theories and show the explicit correspondence.

As this work was being prepared for publication, we became aware of
related work by Ekstrand \cite{Ekstrand} which develops the \v{C}ech
and de Rham pictures in detail. The work of Freund and Nepomechie \cite{FN}
has also been brought to our attention, in which the relationship 
between $(p+1)$-forms over $U(1)$ and connections on $p$-loop spaces
was developed.

\section{Transition functions, Connections, and Loops: An overview of gerbes}

We begin with a definition.
Let $X$ be a manifold and $K$ be a Lie group. A $p$-gerbe $\xi$ on $X$ over $K$
is a pair $(U, g)$, where $U_\alpha$ is a good open cover (one whose
intersections are contractible) of $X$, and
$g_{\alpha_1\alpha_2\cdots\alpha_{p+2}}$ is a collection of functions 
$U_{\alpha_1\cdots\alpha_{p+2}} \equiv U_{\alpha_1}\cap\cdots\cap
U_{\alpha_{p+2}}\rightarrow K$ on every $(p+2)$-fold intersection 
satisfying the {\em inversion condition}
\begin{equation}
g_{\alpha_1\cdots\alpha_i\cdots\alpha_j\cdots\alpha_{p+2}} = 
g_{\alpha_1\cdots\alpha_j\cdots\alpha_i\cdots\alpha_{p+2}}^{-1}
\label{eq:invcond}
\end{equation}
and the {\em cocycle condition} on $(p+3)$-fold intersections

\be
\left(\delta g\right)_{\alpha_1\cdots\alpha_{p+3}} = g_{\alpha_2\cdots\alpha_{p+3}} g_{\alpha_1\alpha_3\cdots
\alpha_{p+3}}^{-1} g_{\alpha_1\alpha_2\alpha_4\cdots\alpha_{p+3}}
\cdots g_{\alpha_1\cdots\alpha_{p+2}}^{(-1)\cdot p} = 1\ .
\label{eq:cocyccond}
\end{equation}

For $p=0$, this reduces to the definition of a $K$-bundle. The definition
(\ref{eq:invcond},\ref{eq:cocyccond})
is based on transition functions and so is somewhat hard to visualize; for
one thing, for $p>0$ a gerbe is not a manifold. (As is the total space of a
bundle) Later on we will see that gerbes nonetheless have a well-defined
notion of section, and in fact a $p$-gerbe is a sheaf of $(p-1)$-gerbes.

We will denote the set of all $p$-gerbes on a given manifold and group by 
$G_p(X,K)$. For consistency in our recursive definitions, we will also
denote by $G_{-1}(X,K)$ the set $C(X,K)$ of continuous functions from $X$
to $K$, and by $G_{-2}(X,K)$ the group $K$ itself.  Since $g$ is a 
$(p+2)$-cocycle, a gerbe $\xi\in G_p(X,K)$ is naturally topologically
classified by the \v{C}ech cohomology group $H^{p+2}(\xi)\equiv H^{p+2}(g\in
C^{p+2}(X,K))$. This is clearly invariant under continuous deformations
(homomorphisms) of $\xi$. Similarly we can naturally define pullbacks
$\omega^\star\xi$ of a gerbe to a submanifold $\omega\subset X$ and 
tensor products $\xi\otimes \xi'$ of gerbes. This construction is the
\v{C}ech picture of gerbes. 

\smallskip

We now define the de Rham (connection) picture. Define the
$\alg K$-valued $(p+2)$-cochain
\begin{equation}
A^{(0)}_{\alpha_1\cdots\alpha_{p+2}} = \log g_{\alpha_1\cdots\alpha_{p+2}}\ .
\end{equation}

Since $g$ is a cocycle, we know that $\delta A^{(0)}=g^{-1}\delta g=1$, and so 
$\delta dA^{(0)} = 0$. This means that
using the Poincar\'e lemma  we can define a 1-form valued $(p+1)$-cochain
$A^{(1)}_{\alpha_1\cdots\alpha_{p+1}}$ satisfying
\begin{equation}
\left(\delta A^{(1)}\right)_{\alpha_1\cdots\alpha_{p+2}} - d A^{(0)}_{\alpha_1
\cdots\alpha_{p+2}} = 0
\end{equation}
on every $(p+2)$-fold intersection. Since $\delta d A^{(1)} = ddA^{(0)} = 0$,
we can repeat this process,
defining a sequence of $n$-form valued $(p-n+2)$-cochains $A^{(n)}$ by
\begin{equation}
\delta A^{(n+1)} - d A^{(n)} =0\ .
\label{eq:itrel}
\end{equation}
Such an iterated use of the Poincar\'e lemma is simply the standard 
relationship of \v{C}ech to de Rham cohomology. This sequence ends when we 
define the $(p+1)$-form $A^{(p+1)}_\alpha$ on every open set
$U_\alpha$, which by the Poincar\'e lemma satisfies
\begin{equation}
dA^{(p+1)}_\alpha = \left.A^{(p+2)}\right|_{U_\alpha}\ .
\end{equation}
$H\equiv A^{(p+2)}$ is a globally defined $(p+2)$-form which is the 
noncovariant field strength of the gerbe.\footnote{We have here used 
partial derivatives, and so this field strength is not the 
one ordinarily used in physics for $K$ non-abelian. In particular, it will
not have the usual gauge invariance. Below we will define a covariant field
strength which remedies this.} $B\equiv A^{(p+1)}$ is its $(p+1)$-form 
connection, which is defined on each coordinate patch.

This process can be reversed, as well; if we are given a closed $(p+2)$-form
$H$ and a contractible cover $U_\alpha$, then by the Poincar\'e lemma
there are $(p+1)$-forms $B$ on each $U_\alpha$ such that 
$\left.H\right|_{U_\alpha}=dB_\alpha$. Then on any $U_{\alpha\beta}$,
$\delta dB = dB_\alpha-dB_\beta = 0$, and so we can define $p$-forms
$A^{(p)}$ such that $\delta B = dA^{(p)}$, and so forth until we come again to
the $0$-forms $A^{(0)}$. Then if $H$ defines an integral class in
$H^{p+2}_{dR}(X,K)$ the exponential  of $A^{(0)}$ is well-defined and so
there are cochains $g=exp\ A^{(0)}$, which therefore form a gerbe. So
to each closed $(p+2)$-form defining an integral class corresponds a
$p$-gerbe. This alows us to pass freely between the \v{C}ech and de Rham
pictures.

There is however an ambiguity in the descending construction. While
we know that $H=dB$, it is possible to shift $B$ by any closed $(p+1)$-form
and maintain this. Therefore $p$-gerbes have a gauge symmetry generated
by a $p$-form:
\begin{equation}
B \rightarrow B + d\xi^{(p)}\ .
\label{eq:gaugesym}
\end{equation}
Similarly there are lower gauge symmetries for each $A^{(n)}$, generated
by $(n-1)$-forms. Once these gauge symmetries are equivalenced out, we
find that

\mydef{Theorem 2.1}{The set of $p$-gerbes is given by the set of 
closed $(p+2)$-forms (field strengths) with integral de Rham 
cohomology class in $H^{p+2}$.}

The gauge symmetry (\ref{eq:gaugesym}) is familiar from the NS $B$-field
in string theory. By this theorem, we can interpret this field as a
connection on a 1-gerbe. This agrees with the result of \cite{Kalk} 
that gerbes describe the $B$-field in massive IIA supergravity. We will
see the geometric interpretation of this relationship below.

\smallskip

At first it may seem odd to define a connection which is not associated
with an obvious covariant derivative. A way to define such a derivative
is suggested by the
result of \cite{Brylinski} that 1-gerbes are equivalent to $K$-bundles over
the loop space $\Omega X$,\footnote{Defined to be the set of embeddings of
$S^1\rightarrow X$, modulo reparametrizations of the $S^1$.} and by a 
theorem due to Getzler, Jones and Petrack \cite{Loops} that the set of 
$k$-forms on $\Omega X$ is isomorphic to the set of 1-cochains of $k$-forms 
on $X$. 

Let us begin with this theorem. If we iterate it, defining the
$p$th loop space by $\Omega^p X\equiv \Omega \Omega^{p-1}X$, $k$-forms
on $\Omega^p X$ are isomorphic to $p$-cochains of $k$-forms on $X$, which
(by the usual exchange of \v{C}ech and spacetime indices) are $(k+p)$-forms
on $X$. Then it is natural to suspect that the $(p+1)$-form $B$ defined
on each $U_\alpha$ can be interpreted as a $1$-form on the loop space
$\Omega^p X$. Using this, we could define a natural action of a $p$-gerbe
on $\Omega^p X$ by means of a covariant derivative
\begin{equation}
\nabla = d + B
\label{eq:covderiv}
\end{equation}
which describes how functions $f(\omega):\Omega^p X\rightarrow K$ transform 
under infinitesimal deformations $\omega\rightarrow\omega+\delta \omega$,
where $\delta\omega\sim \omega\times [0,1]$;
\begin{equation}
f(\omega)\rightarrow f(\omega+\delta\omega) = \exp [\delta\omega\cdot\nabla]
f(\omega)\ .
\end{equation}
The dot product of the $(p+1)$-cochain $\delta\omega$ and $B$ is given by
the usual de Rham product
\begin{equation}
\delta\omega\cdot B = \int_{\delta\omega} B\ .
\end{equation}
The curvature of this covariant derivative is $[\nabla,\nabla]$, a $2$-form
on $\Omega^pX$ which is therefore a $(p+2)$-form on $X$. This is the covariant
generalization of our ordinary field strength. Note that this definition
is meaningful even when $K$ is non-Abelian, and so gives a natural way to 
define Bianchi identities for higher gerbes. However,
although covariant
derivatives can relate the connection to the field strength, there is no
natural way to define the lower forms $A^{(n)}$ in this way, so one can
only pass from the loop picture to the \v{C}ech picture in terms of 
partial derivatives. (The noncovariant field strength is, however, still
defined and useful in the non-Abelian case)

The one subtlety that might obstruct the definition (\ref{eq:covderiv}) is
that a given loop
$\omega$, or its variation $\delta\omega$, may overlap multiple $U_\alpha$
and so no $B$ could be defined on the entire loop. To show that this is
not the case, we will need a result from sections 3 and 4 that gerbes with 
trivial $H^{p+2}$ have a global section and so are equivalent to $(p-1)$-gerbes
on the space. Since 
$\delta\omega$ is $(p+1)$-dimensional, $H^{p+2}(\delta\omega)=0$, and so
the pullback $\delta\omega^\star\xi$ for any $\xi\in G_p(X,K)$ is trivial.
This means that the restriction of $\xi$ to $\delta\omega$ has a global
section, and so $\xi\in G_{p-1}(\delta\omega, K)$. Therefore the
covariant derivative (\ref{eq:covderiv}) can be defined for any gerbe over
the appropriate loop space. Similarly, any connection on $\Omega^p X$ 
can be converted to a collection of $(p+1)$-forms on every open set of $X$,
which by theorem 2.1 defines a $p$-gerbe. We therefore have a
well-defined ``loop picture'' of gerbes, and

\mydef{Proposition 2.2}{$G_p(X,K)\cong G_0(\Omega^pX, K)$. ($p$-gerbes on $X$
are bundles on $\Omega^p X$)}

Since these bundles have a connection, we may interpret this to say that
$p$-gerbes implement gauge symmetries on $p$-fold loop spaces in the same
way that bundles implement gauge symmetries on points. Combining this with
the de Rham picture, this means that $(p+1)$-form connections can 
be interpreted as connections on the space of $p$-loops.

We have not, in this discussion, used the fact that the $\omega$ are actually
loops; we may naturally consider what would happen if instead $\omega\in M^pX$,
the space of smooth $p$-manifolds smoothly embedded in $X$. (This is 
the $p$-dimensional analogue of the unfixed path space) One would expect that
the type of gerbe needed to implement gauge transformations on a manifold
$\omega$ should not change under small deformations of $\omega$ such as
``smearing'' over an interval. Specifically, one expects that if 
$\omega\in M^pX$ is contractible 
to $\eta\in M^qX$, with $q<p$, then a $q$-gerbe should suffice to define 
gauge symmetries on $\omega$. This can easily be shown.
Let $\xi\in G_p(X,K)$; then $\delta\omega^\star\xi$ is trivial, so
$\xi\in G_{p-1}(\delta\omega,K)$. Since $\delta\omega\sim\omega\times[0,1]
\sim\eta\times[0,1]\sim\delta\eta$, we know that 
$\xi\sim\xi'\in G_{p-1}(\delta\eta,K)$. But since $\dim\delta\eta=q+1$,
$H^k(\xi')=0$ for $k>q+1$, and since cohomology classes are invariant
under homomorphism, $H^k(\xi)=0$ for $k>q+1$. Therefore 
$\xi\in G_{q-1}(\delta\omega,K$), and so

\mydef{Theorem 2.3}{Let $M^p_qX=\{\omega\in M^pX: \inf\{\dim \eta:
\eta\sim\omega\} = q\}$. Then every element of $G_q(X,K)$ defines a
connection on $M^p_qX$.

Equivalently, decompose $\bigoplus_p M^p X=\bigoplus_{p,q}
M^p_q X = \bigoplus_q M_q X$, where $M_q = \bigoplus_p M^p_q X$ is the
set of all submanifolds of $X$ which can be contracted down to $q$
dimensions. Then $G_q(X,K)\cong G_0(M_qX, K)$.}

That is, $q$-gerbes define connections on the space of curves homomorphic
to $q$-loops in $X$.\footnote{This generalizes the result of \cite{FN},
which in our language states that $G_q(X,U(1))\cong G_0(M^qX, U(1))$.}
This means that connections on the space of open
strings (embeddings of $[0,1]\rightarrow X$, where $X$ is spacetime) 
take values in $G_0(X,K)$, but connections on closed strings take values
in $G_1(X,K)$ since circles cannot be contracted to a point. We can 
therefore also physically interpret the NS tensor field $B^{\mu\nu}$ as
a connection on a 1-gerbe which implements a $U(1)$ gauge symmetry on the
space of closed strings. This gauge symmetry is identical to the symmetry
which transforms the vector field $A^\mu$ in open string theories in the
absence of background $D$-branes. 

We now have three pictures of gerbes: a \v{C}ech picture, given by open
covers and transition functions; a de Rham picture, given by a $(p+2)$-form
field strength with $(p+1)$-form connections and $p$-form gauge symmetries;
and a loop picture, with bundles on spaces of $p$-loops. It will be
useful for us to introduce still a fourth picture, which will describe
gerbes in terms of their local (section) structure. This picture will both
give additional intuition as to the nature of gerbes and aid in calculations,
especially in the definition of a $K$-theory of gerbes.

\section{Sections of gerbes}

In this section we will restrict ourselves to the case of $K$ abelian.
We wish to determine what a gerbe looks like ``locally,'' i.e. the 
analogue for gerbes of sections of bundles. To do this we will first
define an auxilliary structure called a pregerbe, which is identical
to a gerbe except it does not satisfy the cocycle condition 
(\ref{eq:cocyccond}). Instead we define the variation $\delta\xi$ of 
a pregerbe $\xi$ to be the set of coboundaries of the transition functions
of the pregerbe;
\begin{equation}
h_{\alpha_1\cdots\alpha_{p+3}} \equiv \left(\delta g\right)_{\alpha_1
\cdots\alpha_{p+3}}\ .
\end{equation}
A pregerbe is a gerbe if all elements of $\delta\xi$ are unity.
By the Poincar\'e lemma, $\delta h=\delta\delta g=1$, and so the
variation of any pregerbe is a gerbe. We denote the class of $p$-pregerbes
by $PG_p(X,K)$.

We also define a notion of equivalence for two pregerbes on the
same manifold. For $\xi=(U, g)$ and $\xi'=(U',g')$, we define the
mutual refinement $U\cap U'$ of the two covers to be the set of all 
intersections of elements of $U$ with elements of $U'$. Clearly both $\xi$ and
$\xi'$ have a natural extension to this mutual refinement. Then we
say that $\xi\cong\xi'$ if $\delta\xi=\delta\xi'$ on each $(p+3)$-fold
intersection in $U\cap U'$. (i.e., if their variations define the same
gerbe)

We begin by proving a simple but useful lemma. We define the difference of
two pregerbes $\xi=(U,g)$ and $\chi=(U',g')$ to be 
\begin{equation}
\chi-\xi =( U\cap U', g'_{\alpha_1\cdots \alpha_{p+2}} g^{-1}_{\alpha_1\cdots
\alpha_{p+2}})\ .
\end{equation}
Then

\mydef{Lemma 3.1}{$\delta (\chi-\xi) = 1$ iff $\chi \cong \xi$. (The difference
of two equivalent pregerbes is a gerbe)}

{\bf Proof.} This follows from direct evaluation of the variation. On a
$(p+3)$-fold intersection $U_{\alpha_1\cdots\alpha_{p+3}}$,
\bea
\delta (\chi-\xi) &=& (g'g^{-1})_{\alpha_2\cdots\alpha_{p+3}}
(g'g^{-1})_{\alpha_1\alpha_3\cdots\alpha_{p+3}}^{-1} \cdots
(g'g^{-1})_{\alpha_1\cdots\alpha_{p+2}}^{(-1)^p} \nn \\
&=& g'_{\alpha_2\cdots\alpha_{p+3}}g'^{-1}_{\alpha_1\alpha_2\cdots\alpha_{p+3}}
\cdots g'^{(-1)^p}_{\alpha_1\cdots\alpha_{p+2}}
g^{(-1)^{(p+1)}}_{\alpha_1\cdots\alpha_{p+2}} g_{\alpha_1\alpha_3\cdots
\alpha_{p+3}}\cdots g^{-1}_{\alpha_2\cdots\alpha_{p+3}} \nn \\
&=& \delta g'_{\alpha_1\cdots\alpha_{p+3}} (\delta g_{\alpha_1\cdots\alpha_{p+3}}
)^{-1} \nn
\eea
Which is equal to unity iff the two variations are equal on all intersections.

This associates a unique $p$-gerbe with each equivalence class of $p$-pregerbes.
Likewise every gerbe can be written as the variation of some pregerbe; thus

\mydef{Lemma 3.2}{The set of equivalence  classes in $PG_p$ is isomorphic to
$G_p$.}

\medskip

We can describe the local structure of gerbes in terms of pregerbes. We
say that a local trivialization of a gerbe $\xi=(U,g)$ on a $(p+1)$-fold
intersection $U_{\alpha_1\cdots\alpha_{p+1}}$ is a realization of the 
cocycle $g$ as a coboundary, i.e. 
a collection of functions $f_{\alpha_1\cdots\alpha_{p+1}}$ 
such that 
\begin{equation}
g_{\alpha_1\cdots\alpha_{p+2}}=(\delta f)_{\alpha_1\cdots\alpha_{p+2}}
=f_{\alpha_2\cdots\alpha_{p+2}}f_{\alpha_1\alpha_3\cdots\alpha_{p+2}}^{-1}\cdots
f_{\alpha_1\cdots\alpha_{p+1}}^{(-1)^{(p-1)}}
\end{equation}
The $f^{(\alpha)}$ are simply a $(p-1)$-pregerbe on
$U_{\alpha_1\cdots\alpha_{p+1}}$ whose variation is $\xi$.  Clearly,
such trivializations are not unique; the set of all local
trivializations of a given $\xi$ is an equivalence class of pregerbes
on $U_{\alpha_1\cdots\alpha_{p+1}}$. A trivialization of $\xi$ is a 
collection of local trivializations on every such intersection; a global
trivialization is a single local trivialization defined simultaneously 
over all of $X$. We will see that the condition for a global trivialization
to exist is that $H^{p+2}(\xi)$ is trivial. 

Trivializations, however, can always be constructed. One trivialization of
particular interest is given by choosing on each $U_\alpha$
\begin{equation}
f^{(\alpha)}_{\beta_1\cdots\beta_{p+1}} \equiv g_{\alpha\beta_1\cdots
\beta_{p+1}}
\label{eq:sectiondef}
\end{equation}
This is then defined on each $(p+1)$-fold intersection, and 
forms a $(p-1)$-pregerbe on each $U_\alpha$, whose variation is the
restriction of the original gerbe to that set. The collection of all such
$f^{(\alpha)}$ forms a trivialization valid on each $U_\alpha$, since
\bea
\left(\delta f^{(\alpha)}\right)_{\beta_1\cdots\beta_{p+2}} &=&
f^{(\alpha)}_{\beta_2\cdots\beta_{p+2}} \cdots 
f^{(\alpha)[(-1)^{(p-1)}]}_{\beta_1\cdots\beta_{p+1}} \nn \\
&=& g_{\alpha\beta_2\cdots \beta_{p+2}}\cdots g_{\alpha\beta_1\cdots
\beta_{p+1}}^{(-1)^{(p-1)}} \nn \\
&=& g_{\beta_1\cdots\beta_{p+2}}\left(\delta g\right)_{\alpha\beta_1
\cdots\beta_{p+2}} \nn \\
&=& g_{\beta_1\cdots\beta_{p+2}}
\eea
We call each $f^{(\alpha)}$ a section of $\xi$ on $U_\alpha$.

This term is justified by showing that the set of such $f^{(\alpha)}$
forms a sheaf of pregerbes, which (since each $f^{(\alpha)}$ is a 
representative of an equivalence class of pregerbes)
makes the set of local trivializations of $\xi$ into a sheaf of
equivalence classes of pregerbes. To
do this, we note that on each $U_\alpha$ we have defined a collection
of functions $f^{(\alpha)}_{\beta_1\cdots\beta_{p+1}}$ which map the
intersection $U_\alpha\cap U_{\beta_1\cdots\beta_{p+1}}$ to $K$. If
the $f^{(\alpha)}$ form a sheaf, there must be transition
functions for each of these functions on intersections 
$U_\alpha\cap U_{\alpha'}$. These follow from the cocycle condition
on $\xi$;
\begin{equation}
\left(\delta g\right)_{\alpha\alpha'\beta_1\cdots\beta_{p+1}} = g_{\alpha'\beta_1\cdots\beta_{p+1}} g_{\alpha\beta_1\cdots
\beta_{p+1}}^{-1} g_{\alpha\alpha'\beta_2\cdots\beta_{p+1}}
\cdots g_{\alpha\alpha'\beta_1\cdots\beta_p}^{(-1)^p} = 1
\end{equation}
and so
\begin{equation}
(f^{(\alpha)})_{\beta_1\cdots\beta_{p+1}} = (\phi_{\alpha
\alpha'})_{\beta_1\cdots\beta_{p+1}} (f^{(\alpha')})_{\beta_1
\cdots\beta_{p+1}}
\end{equation}
where
\begin{equation}
\phi_{\alpha\alpha'} = g_{\alpha\alpha'\beta_2\cdots\beta_{p+1}}
g_{\alpha\alpha'\beta_1\beta_3\cdots\beta_{p+1}}^{(-1)}
\cdots g_{\alpha\alpha'\beta_1\cdots\beta_p}^{(-1)^p}\ .
\end{equation}
The $\phi_{\alpha\alpha'}$ clearly satisfy the inversion condition 
$\phi_{\alpha\alpha'}\phi_{\alpha'\alpha}=1$; they also satisfy the 
cocycle condition
\begin{equation}
\phi_{\alpha_1\alpha_2}\phi_{\alpha_2\alpha_3}\phi_{\alpha_3\alpha_1} = 
f^{(\alpha_1)} f^{(\alpha_2)}{}^{-1} f^{(\alpha_2)} f^{(\alpha_3)}{}^{-1} 
f^{(\alpha_3)} f^{(\alpha_1)}{}^{-1} = 1
\end{equation}
and so they indeed are the transition functions on a sheaf. Therefore
(since on each $U_\alpha$ this trivialization is a representative of
the equivalence class of all local trivializations) we see that

\mydef{Lemma 3.3}{$G_p$ is isomorphic to the set of equivalence classes
of sheaves of $PG_{p-1}$.}

It then follows from lemmas 3.2 and 3.3 that

\mydef{Theorem 3.4}{The set of $p$--gerbes is isomorphic to the set of
sheaves of $(p-1)$--gerbes.}

{\bf Proof.} Each element of $G_p$ is isomorphic to a equivalence class of
sheaves of $PG_{p-1}$, which is isomorphic to a sheaf of equivalence  
classes of $PG_{p-1}$, which is a sheaf of $(p-1)$--gerbes.

This allows us to think of gerbes as sheaves of lower
gerbes. For $p=0$ this is trivial, simply stating that 0-gerbes are
sheaves whose sections are continuous functions. For $p=1$ we can
consider the example given in \cite{Hitchin} of the gerbe of spin
structures on a space which admits a global $SO$ structure but 
not a $Spin_c$ structure. In such a
case it is natural to cover the space with open sets, on each of 
which it is possible to define a $Spin_c$ structure, and 
define transition functions between the structures. Since each structure
is itself a line bundle (specifically, an $S^1$-bundle) this construction
is a 1-gerbe whose sections are local $Spin_c$ structures. The 
cohomology group $H^{p+2}$ associated with this gerbe is essentially the mod 2
reduction of the second Steifel-Whitney class $w_2(P)$ (where $P$ is 
the $SO$ bundle) whose triviality implies that a $Spin_c$ structure
can be globally defined. In this case (in the language of theorem 3.4)
the gerbe would be topologically trivial and so has a  global section, in this
case the 0-gerbe (bundle) of $Spin_c$ structure.

\medskip

We can take this construction slightly farther by noting that $G_p(X,K)$
forms a group. This can be shown by induction. The statement is
clearly true for $p=-1$, using pointwise multiplication. Now if it is
proven for some $p$, then an element of $G_{p+1}(X,K)$ is a sheaf of
groups. We define the product of two sheaves by the pointwise multiplication
of sections; i.e., if $\xi=(U_\alpha, s_\alpha)$ and 
$\xi'=(U'_\alpha,s'_\alpha)$, then $\xi\xi'=(U\cap U', (ss')_\alpha)$. The
transition functions for this sheaf are $\phi_{\alpha\beta}=s_\alpha s'_\alpha
s'^{-1}_\beta s^{-1}_\beta$. This clearly satisfies the group axioms, with
the trivial sheaf acting as identity. Therefore an element of $G_p$ is 
actually a bundle with sections in $G_{p-1}$, and so structural group
$G_{p-2}$. Thus

\mydef{Corollary 3.5}{$G_p(X,K)\cong G_0(X,G_{p-2}(X,K))$.}

This generalizes the theorem \cite{Hitchin, Chatterjee} that Abelian
1-gerbes can be described as bundles of bundles. While we have used
the Abelian property in deriving this result, we believe that a very
similar result should hold in the non-Abelian case. In such a case,
we know by this argument that $p$-gerbes are sheaves of equivalence
classes of $(p-1)$-pregerbes, but not that these themselves form
$(p-1)$-gerbes. The definition of a section structure continues to
hold in this case. In both cases, it is clear that the existence of a
global section of the sheaf (some $U_\alpha=X$) is equivalent to the
existence of such for the gerbe to which it is associated, and so
the conditions for their topological triviality must be equivalent.
This is essentially the result used in the proof of proposition 2.2;
we will develop it in a slightly more detailed form in section 4
as well.

We can also relate the sheaf picture to the de Rham picture. The
connection on the sheaf associated to a $p$-gerbe is the $G_{p-2}$-valued
one-form $\delta\log f^{(\alpha)}$, which by construction is equal to 
the one-form $A^{(1)}$ defined
earlier. If we transform the \v{C}ech indices of $f^{(\alpha)}_{\beta_1\cdots
\beta_{p+1}}$ to spacetime indices as before,
the sheaf connection is then equal to the gerbe connection, with one
index of the gerbe connection corresponding to the one-form index of 
the sheaf connection, and the rest interpretable as internal indices.
The relation to the loop picture is less clear, but can be found
by going through the de Rham construction.

\section{$K$--theory of gerbes}

The fact that gerbes are also sheaves suggests that they should have a 
natural $K$-theory. A natural choice is to define the $K$-theory of
gerbes to simply be that of the associated sheaves; we will show that
this is the same $K$-theory as one would get by directly defining the
Whitney sum of gerbes. This $K$-theory then will classify sources of
$B$-field charge (for example) in the same way that the usual $K$-theory
of bundles classifies 1-form charges.

So in this section we will do the following: First, we will show
that the gerbe Whitney sum agrees with the Whitney sum of sheaves related
to the gerbes. We will use this to define the ``topological'' $K$-theory
of gerbes (analogous to the topological $K$-theory of bundles) and show
that it behaves like one would expect a $K$-theory to behave. We will then
demonstrate the analogue of the Serre-Swan theorem, which for bundles
relates their topological $K$-theory to the algebraic $K$-theory of 
the ring $C(X,K)$, and for gerbes allows us to relate this topological
$K$-theory to an algebraic $K$-theory of $\alg G_{p-2}$. This will give
us the second recursion relation, which will allow us to make 
explicit calculations of $K^0$.

\bigskip

We begin with the Whitney sum of sheaves. To each $p$-gerbe $\xi$ is
associated a sheaf whose sections are the
$f^{(\alpha)}$. The Whitney sum of the two sheaves associated to
$\xi$ and $\xi'$ then has sections $f^{(\alpha)}\oplus f^{'(\alpha)}$
on each set in their mutual refinement. The addition $\oplus$ is 
simply the direct sum of two $K$-representations. By (\ref{eq:sectiondef}),
this means that the ``sheaflike'' Whitney sum of two gerbes is 
another gerbe, with transition functions
\begin{equation}
g_{\alpha\beta_1\cdots\beta_{p+1}}\oplus g'_{\alpha\beta_1\cdots
\beta_{p+1}}\ .
\end{equation}
This is precisely what we would naturally define as the Whitney
sum of two gerbes in the absence of any notion of associated sheaves.
Therefore we can refer to this addition as the Whitney sum of gerbes
without any hesitation.

Since this sum is a Whitney sum of sheaves, though, Swan's theorem
applies, so that for every $\xi$ there is a $\xi'$ such that $\xi\oplus\xi'$
is trivial. (In the sheaf sense, that is that its class in  $H^2(X,G_{p-2})$
is trivial) This triviality means that the sheaf
associated to the sum posesses a global section; but this implies that
the gerbe itself has a global section, and so $\xi\oplus\xi'$ is trivial
in the gerbe sense as well. (That is, its cohomology in $H^{p+2}(X,K)$ is
trivial) This proves that

\mydef{Lemma 4.1}{(Swan's Theorem for gerbes) For every $\xi\in G_p(X,K)$,
there exists a $\xi'\in G_p(X,K)$ such that $\xi\oplus\xi'$ is trivial.}

The Whitney sum therefore gives the set of homomorphism classes of  
$p$-gerbes the structure of a
monoid, just as it does for sheaves. We can therefore define the $K$-group
of $p$-gerbes $K^0[G_p(X,K)]$ to be the enveloping (Grothendieck) group
of this monoid.
By the relationship of
Whitney sums of gerbes to the sums of the associated sheaves, the $K$-group
of gerbes is equal to the $K$-group of sheaves, so $K^0$ commutes with
the isomorphism of Corollary 3.5; i.e.,

\mydef{Lemma 4.2}{$K^0[G_p(X,K)] = K^0[G_0(X,G_{p-2}(X,K))]$.}

This allows us to calculate $K$-groups of gerbes using the technology 
already developed for calculating the same groups for sheaves. It also
means that the usual theorems of $K$-theory -- in particular, the
exact sequences and Bott periodicity -- continue to apply to the $K$-theory
of gerbes.

There is one particular theorem which it is worth examining in this
case, namely the Serre-Swan theorem. This theorem ordinarily states that
the topological $K$-theory of fiber bundles (the construction described
above for $p=0$) is isomorphic to the algebraic $K$-theory $K_0$ of the
module $\Gamma$ of sections of bundles,\footnote{The $K$-theory of a $C^\star$-algebra
such as $C(X,K)$ is defined (for algebras posessing a unit) to be the 
enveloping group of the monoid of homomorphism classes of projection 
operators in the algebra under a Whitney sum. This algebraic $K$-theory 
generalizes the ordinary topological $K$-theory of bundles, which is
algebraically the $K$-theory of commutative algebras.
The problem of non-unital $C^\star$-algebras is
analogous to that of bundles on noncompact spaces, and is resolved by
taking a unital extension of the algebra and then modding out its 
contribution to the $K$-group. The analogous procedure for topological
$K$-theory is to move to the one-point compactification of the space, e.g.
$K^0(\R^n)\equiv K^0(S^n)$. An accessible introduction to algebraic $K$-theory
is given in \cite{WeggeOlsen}.} i.e.
\begin{equation}
K^0[G_0(X,K)] \cong K_0[\Gamma[G_0(X,K)]] = K_0[C(X,K)]\ .
\end{equation}

Using propositon 3.4, this implies that

\smallskip
\mydef{Proposition 4.3}{$K^0[G_p(X,K)] \cong K_0[\Gamma[G_p(X,K)]] =
K_0[G_{p-1}(X,K)]$,}
where the quantity on the left-hand side is the topological $K$-theory
of gerbes defined above, and the quantity on the right-hand side is the
algebraic $K$-theory of the group of $(p-1)$-gerbes defined in section
3. This and corollary 3.5 are our recursion relations. They can naturally
be used to compute $K$-groups; for instance, 
\bea
K^0[G_1(X,K)] &=& K^0[G_0(X,G_{-1}(X,K))] \nn \\
&=& K_0[G_{-1}(X,C(X,K))] \nn \\
&=& K_0[C(X,C(X,K))] \nn \\
&=& K_0[C(X^2,K)]\ .
\eea

This can be used along with the ordinary Sen construction \cite{Sen1}-\cite{Sen5}
to determine the allowable types of NS $B$-field charge for branes in 
type II string theory. The process works identically to the $K$-theory
classification of 1-form charges, \cite{Witten, Horava, Hori} now using 1-gerbes
and their Whitney sums. In type IIB, one describes $p$-branes as defects
in D9-D$\bar 9$ pairs. Then as for bundles, (all of the same reasoning
applies) the $B$-field charge of a $p$-brane takes values in
\be
K^0[G_1(\R^{10-p},\C)]=K_0[C(\R^{20-2p},\C)]=\Z\ ,
\eeee
where the latter is a standard result of topological $K$-theory.

In type IIA, there is the additional subtlety that 9-branes are not
stable and so the simplest version of the Sen construction does not
suffice. In this case, analogy with the bundle case suggests that the
solution is to take a higher $K$-group $K^{-1}$. This group is defined
for bundles as the group of pairs $(E,\alpha)$, where $E$ is a bundle
and $\alpha$ is an automorphism of $E$, with addition rule
$(E,\alpha)+(F,\beta) = (E\oplus F, \alpha\oplus\beta)$ and modulo
the equivalence $(E,\alpha)\sim (F,\beta)$ if there exist $(E',\alpha')$
and $(F',\beta')$ such that $\alpha'$ and $\beta'$ are homomorphic 
to the identity automorphism and $(E,\alpha)+(E',\alpha')\cong (F,\beta)+
(F',\beta')$. We take the same definition for gerbes. As for $K^0$,
this definition is insensitive to whether we use the gerbe or the
sheaf Whitney sum, and the equivalent of lemma 4.2 applies as well.
Since this is equivalent to a $K$-group of sheaves, the analogue of
proposition 4.3 is valid as well; in this case, it is
\be
K^{-1}[G_p(X,K)] \cong K_1[\Gamma[G_p(X,K)]] = K_1[G_{p-1}(X,K)]\ .
\eeee
The group $K_1$ is another algebraic $K$-group.

We will not attempt to show whether $K^{-1}$ classifies gerbe charges
in type IIA theory as it does for bundles, but this is a reasonable
expectation. If indeed it does, then we may calculate
\bea
K^{-1}[G_1(X,\C)] &=& K^{-1}[G_0(X,C(X,\C))] \nn \\
&=& K_1[C(X^2,\C)] \nn \\
&=& K^{-1}[G_0(X^2,\C)] \nn 
\eea
By Bott periodicity, this is equal to $K^0[G_0(SX^2,\C)]$, where
the suspension $SY$ of a manifold $Y$ is defined to be the one-point
compactification of $Y\times[0,1]$. In particular, for $X=\R^{10-p}$,
$SX^2=S^{21-2p}$, and so the $B$-field takes values in
$K^0[G_0(S^{21-2p},\C)] = \Z$.

 Therefore in both type II theories, all
branes (both stable and unstable) can carry an integral NS $B$-field
charge. This is not surprising since all such branes couple to the
fundamental string, but is a good check on our picture of gerbes.

One should note that by construction, the modules of gerbes are 
commutative, and so in a sense all of the $K$-groups one finds for 
gerbes are the same as those found for bundles. But this should come
as no surprise, since as we have seen gerbes are themselves very
special bundles. 

\vskip 0.25in

Gerbes are therefore prone to arise under a wide variety of
circumstances. By theorem 2.1, any higher-form connection (a higher-form
field with appropriate gauge symmetries and transition functions,
or equivalently a higher-form field strength) leads to a gerbe; 
by theorem 3.4, whenever there is a topological obstruction to
forming an (Abelian) bundle we have a gerbe. Finally, by theorem 2.3
any gauge symmetry where the transforming objects are extended objects
is naturally described by a gerbe. 

It is interesting that the converses of these statements are true as
well. In particular, the presence of a higher-form connection implies
a gauge symmetry realized on extended objects in the theory. Consider,
for example, the case $p=1$, where we have a 2-form $B^{\mu\nu}$ with
field strength $H=\nabla B$. Let us restrict our attention to the
case of $K$ abelian so that we need not concern ourselves with the
distinction between $\nabla$ and $d$. Then our action is likely to
contain terms such as $H\wedge\star H$ (for a Yang-Mills-like theory)
or $B\wedge H$. (For a Chern-Simons-like theory) Using (\ref{eq:itrel}),
we can write at least formally $B=\delta^{-1}dA$, where $A$ is the
$\alg K$-valued 1-form connection. The inverse coboundary operator $\delta^{-1}$
is clearly nonlocal; it effectively integrates over a 1-cochain. 
Therefore we can consider our connection $B$ to be the integral
of a 2-form over a 1-dimensional internal space. This is of course 
consistent with our loop picture, since $B$ is a connection over $M_1X$. 

For $p=1$, it is also straightforward to evaluate $M_1X$; by definition,
it is the space of submanifolds homomorphic to dimension-1 submanifolds
which are not homomorphic to dimension-0 submanifolds, i.e. points, and
so $M_1X$ is simply the space of submanifolds homomorphic to loops in $X$.
Therefore this theory may be reinterpreted as a theory of 1-forms taking
values in $\alg K\times \Omega^1 X$, and the extended objects found in
our theory are closed strings. Similar arguments can be made for higher
$p$; for example, for $p=2$ and $X$ compact, $M_2X$ is the space of 
(submanifolds homomorphic to) Riemann surfaces in $X$, while for $X$ 
noncompact $M_2X$ also includes the family of infinite membranes. This
result agrees with the known relationship of $p$-forms and extended 
objects in $M$-theory. 

It is therefore natural to consider $p$-gerbes to be the generalization
of bundles relevant to theories which have higher forms and extended
objects. Using the geometric constructions and the $K$-theory defined
above, these can be treated on a reasonable physical footing; they 
posess conserved charges, covariant field strengths, and gauge symmetries.
However, several important issues, notably the definition of the
lower-form connections and the sheaf picture in the non-Abelian case,
still must be resolved.

\medskip

\centerline{\sc Acknowledgements}

The author wishes to thank Ralph Cohen, Robbert Dijkgraaf, Edi Halyo, 
and John McGreevy for useful conversations and comments. This work was 
partially supported by an NSF graduate research fellowship.

\end{document}